\documentclass[letterpaper,notoc]{CECS}
\usepackage{amsfonts}
\title{\bf Canonical sectors of five-dimensional\\ Chern-Simons theories}
\author{Olivera Mi\v{s}kovi\'{c}$^{\ast ,\dagger }$, Ricardo Troncoso$%
^{\ast} $ and Jorge Zanelli$^{\ast} \medskip \medskip $ \\
$^{\ast }${\small \emph{Centro de Estudios Cient\'{\i}ficos (CECS), Casilla
1469, Valdivia, Chile}.}\\
$^{\dagger }${\small \emph{Departamento de F\'{\i}sica, P. Universidad Cat%
\'{o}lica de Chile, Casilla 306, Santiago 22, Chile}.}}

\preprint{{\tiny CECS-PHY-04/02} }

\abstract{The dynamics of five-dimensional Chern-Simons theories is analyzed.
These theories are characterized by intricate self couplings which give rise to
dynamical features not present in standard theories. As a consequence, Dirac's
canonical formalism cannot be directly applied due to the presence of
degeneracies of the symplectic form and irregularities of the constraints on
some surfaces of phase space, obscuring the dynamical content of these
theories. Here we identify conditions that define sectors where the canonical
formalism can be applied for a class of non-Abelian Chern-Simons theories,
including supergravity. A family of solutions satisfying the canonical
requirements is explicitly found. The splitting between first and second class
constraints is performed around these backgrounds, allowing the construction of
the charge algebra, including its central extension.}

\begin{document}

\section{Introduction}

The best known gauge theories whose dynamical field is a connection on a
fiber bundle are described by Yang-Mills and Chern-Simons (CS) actions.
Three-dimensional CS theories are topological and also provide descriptions
for gravitation and supergravity \cite{Achucarro-Townsend,Witten'88}. There
also exist gravity theories in higher odd dimensions described in terms of
CS actions \cite{Chamseddine1,Troncoso-Zanelli'00}. For negative
cosmological constant, in five dimensions the locally supersymmetric
extension was found in \cite{Chamseddine2}, and for higher odd dimensions in
\cite{Troncoso-Zanelli'98,Troncoso-Zanelli'99,Troncoso-Zanelli'99a}. For
vanishing cosmological constant supergravity theories sharing this geometric
structure have also been constructed in \cite%
{Banados-Troncoso-Zanelli,HOT,Hassaine-Troncoso-Zanelli}. However, this
elegant geometrical setting leads to a rich and quite complex dynamics.
Indeed, for the purely gravitational sector, the Lagrangian in $D=5$
dimensions, apart from the Einstein-Hilbert Lagrangian, also contains the
Gauss-Bonnet term which is quadratic in the curvature, while for $D\geq 7 $
additional terms with higher powers of the curvature and explicitly
involving torsion are also required.

CS theories for $D\geq 5$ have been studied in different contexts (see \emph{%
e. g.}, \cite{Floreanini-Percacci,Floreanini-Percacci-Rajaraman,Nair-Schiff}%
), and are not necessarily topological theories but can contain propagating
degrees of freedom \cite{Banados-Garay-Henneaux}. Their dynamical structure
depends on the location in phase space, and can drastically change from
purely topological sectors to others with different numbers of local degrees
of freedom. Sectors where the number of degrees of freedom is not maximal
are \emph{degenerate} and on them additional local symmetries emerge \cite%
{Saavedra-Troncoso-Zanelli}.

Furthermore, the symmetry generators in CS theories may be functionally
dependent in some regions of phase space. Sectors where this happens are
called \emph{irregular} \cite{Henneaux-Teitelboim-book,Miskovic-Zanelli}.
Around irregular configurations the standard Dirac procedure, required to
identify the physical observables (propagating degrees of freedom, conserved
charges, etc.), is not directly applicable. Furthermore, the naive
linearization of the theory fails to provide a good approximation to the
full theory around irregular backgrounds \cite%
{Chandia-Troncoso-Zanelli,Miskovic}.

Degeneracy and irregularity are two independent conditions which may occur
simultaneously in CS theories, and it is not yet fully understood how to
deal with them. Irregular sectors are also found in the Plebanski theory
\cite{Buffenoir-Henneaux-Noui-Roche}. Although these features are rarely
found in field theories, they naturally arise in fluid dynamics, as in the
description of vortices through the Burgers equation \cite{Burgers}, or in
transonic wave propagation in compressible fluids described by the Chaplygin
and Tricomi equations \cite{Landau-Lifschitz}.

The presence of degenerate and irregular sectors obscures the dynamical
content of these theories, as Dirac's canonical formalism cannot be directly
applied to them. In section 2, for simplicity we consider a non-Abelian CS
theory in five dimensions, which captures the dynamical behavior without
loss of generality. In section 3, we identify conditions that define
canonical sectors, that is, those where the canonical formalism can be
applied, and a family of solutions satisfying the canonical requirements is
explicitly found. In Section 4, the splitting between first and second class
constraints is performed around these backgrounds, allowing the construction
of the charge algebra, including its central extension. Section 5 contains
the discussion and outlook.

\section{Dynamics}

Chern-Simons Lagrangians describe gauge theories for a Lie-algebra-valued
connection $\mathbf{A}=A_{\mu }^{K}\,\mathbf{G}_{K}\,dx^{\mu }$\thinspace\
where $K=1,...,\Delta $, and $\Delta $ is the dimension of the Lie group.
The five-dimensional Chern-Simons form is such that its exterior derivative
is an invariant 6-form,
\begin{equation}
dL=k\,\left\langle \mathbf{F}^{3}\right\rangle =k\,g_{KLM}\,F^{K}\wedge
F^{L}\wedge F^{M},  \label{dL 5}
\end{equation}%
where $\mathbf{F}=d\mathbf{A}+\mathbf{A}\wedge \mathbf{A}=F^{K}\mathbf{G}%
_{K} $ is the field strength 2-form and $k$ is a dimensionless constant.
Here $\langle \cdots \rangle $ stands for a symmetrized\footnote{%
In case of a superalgebra, the standard (anti)symmetrized form is assumed.}
trilinear invariant form, which defines the third rank invariant tensor $%
g_{KLM}=\langle \mathbf{G}_{K}\,,\mathbf{G}_{L}\,,\mathbf{G}_{M}\rangle $.
The action is\footnote{%
Hereafter, wedge products between forms will be omitted.}%
\begin{equation}
I\left[ \mathbf{A}\right] =\int\limits_{M}L(\mathbf{A})=k\int\limits_{M}%
\left\langle \mathbf{AF}^{2}-\frac{1}{2}\,\mathbf{A}^{3}\mathbf{F}+\frac{1}{%
10}\,\mathbf{A}^{5}\right\rangle \,,  \label{I5}
\end{equation}%
where $M$ is a five-dimensional manifold. The field equations read
\begin{equation}
\left\langle \mathbf{F}^{2}\mathbf{G}_{K}\right\rangle =0\,.
\label{Field Eqns. with Brackets}
\end{equation}

\subsection{Warming up with the linearized approximation}

Non-Abelian CS theories are characterized by intricate self couplings which
give rise to dynamical features not present in standard theories. These
non-trivial properties can be captured in the linearized theory in five
dimensions. The action in the linear approximation around a background
solution $\mathbf{\bar{A}}$ has the form $I[\mathbf{\bar{A}}+\mathbf{u}]=I[%
\mathbf{\bar{A}}]+I_{\mbox{{eff}}}[\mathbf{u}]$, where the effective action
is\footnote{%
The five-dimensional manifold is assumed to be topologically $\mathbb{R}%
\otimes \Sigma $, and the coordinates are chosen as $x^{\mu }=(x^{0},x^{i})$%
, where $x^{i}$, with $i=1,\ldots ,4$ correspond to the space-like section $%
\Sigma $.}
\begin{equation}
I_{\mbox{eff}}\left[ \mathbf{u}\right] =3k\int\limits_{M}\left\langle
\mathbf{u\bar{F}}\bar{\nabla}\mathbf{u}\right\rangle =\int d^{5}x\,\left(
\frac{1}{2}\,u_{i}^{K}\bar{\Omega}_{KL}^{ij}\bar{\nabla}%
_{0}u_{j}^{L}-u_{0}^{K}C_{K}-h\right) ,  \label{linearapprox}
\end{equation}%
with the \textquotedblleft potential\textquotedblright\ $h\left( \mathbf{u}%
\right) \equiv 3k\,\varepsilon ^{ijkl}g_{KLM}\,u_{i}^{K}\bar{F}_{0j}^{L}\bar{%
\nabla}_{k}u_{l}^{M}$ and the covariant derivative $\bar{\nabla}\mathbf{u}=d%
\mathbf{u}+\left[ \mathbf{\bar{A}},\mathbf{u}\right] $. The constraint $%
C_{K} $ is given by
\begin{equation}
C_{K}=\bar{\Omega}_{KL}^{ij}\,\bar{\nabla}_{i}u_{j}^{L},
\end{equation}%
and the kinetic term is defined by the symplectic form
\begin{equation}
\bar{\Omega}_{KL}^{ij}\equiv \Omega _{KL}^{ij}(\bar{F})=-3k\,g_{KLM}\,%
\varepsilon ^{ijkl}\bar{F}_{kl}^{M}\,,  \label{symplectic Matrix}
\end{equation}%
which explicitly depends on the curvature.

Thus, the time evolution of the perturbations $u_{i}^{K}$ depends on the
background field strength, $\bar{F}_{kl}^{K}$, and hence the dynamics is
crucially sensitive to the particular background around which it is
explored. Overlooking this issue may lead to the paradoxical situation that
the linearized theory around some backgrounds may seem to have more degrees
of freedom than the fully nonlinear theory \cite{Chandia-Troncoso-Zanelli}.
This is due to the fact that around those backgrounds, the linear
approximation eliminates some constraints from the action.

Since the rank of $\bar{\Omega}$ is not fixed, the number of dynamical
degrees of freedom can change throughout phase space:

\emph{Generic} configurations have maximal rank of $\bar{\Omega}$ and form
an open set in phase space. The theory around this kind of configurations
has maximal number of degrees of freedom \cite{Banados-Garay-Henneaux}.

\emph{Degenerate} configurations are the ones for which the rank of $\bar{%
\Omega}$ is not maximal so that they have additional gauge symmetries, and
thus fewer degrees of freedom.

On the other hand, not only the rank of the symplectic form can vary, but
the linear independence of the constraints $C_{K}$ is not guaranteed either,
and can fail on some backgrounds.

If the constraints $C_{K}$ are independent, the sector is \emph{regular}.
This is the case in all standard theories. However, in CS theories, the
constraints $C_{K}$ can become dependent on some backgrounds, and these
sectors are called \emph{irregular}.

In an irregular configuration, there is always a linear combination of $C$'s
that identically vanishes. Consequently, in the linear approximation the number
of degrees of freedom seems to increase in irregular sectors. However, this is
an illusion induced by using the linear approximation which is no longer valid
in this case. Moreover, the Dirac approach is not directly applicable in
irregular sectors \cite{Dirac}.

Indeed, careful analysis of the full non-linear theory shows that the number
of degrees of freedom cannot increase. This is discussed in the next section.

\subsection{Nonlinear dynamics}

The field equations (\ref{Field Eqns. with Brackets}) can be written as
\begin{equation}
\varepsilon ^{\mu \nu \lambda \rho \sigma }g_{KLM}\,F_{\mu \nu
}^{L}F_{\lambda \rho }^{M}=0\,.  \label{covariant eqs}
\end{equation}%
Therefore, the field equations (\ref{covariant eqs}) split into the
dynamical equations,
\begin{equation}
\Omega _{KL}^{ij}\,F_{0j}^{L}=0\,,  \label{fieldEQ}
\end{equation}%
and the constraints,
\begin{equation}
C_{K}=\frac{1}{4}\,\Omega _{KL}^{ij}\,F_{ij}^{L}\approx 0\,.
\label{constraints}
\end{equation}%
The symplectic matrix $\Omega _{KL}^{ij}(F)$, defined in Eq. (\ref%
{symplectic Matrix}), is a $4\Delta \times 4\Delta $ array with indices $%
\left( _{K}^{i}\right) $ and $\left( _{L}^{j}\right) $ with at least four
zero modes (since $\Omega _{KL}^{ij}F_{jk}^{L}=\delta _{k}^{i}C_{K}\approx 0$%
), corresponding to the spatial diffeomorphisms. The existence of these four
zero modes implies that the rank of $\Omega $ cannot exceed $(4\Delta -4)$
\cite{Banados-Garay-Henneaux}.

As in the linearized approximation, the symplectic matrix is a function of
the field strength $F_{ij}^{K}$, its rank is not necessarily constant
throughout phase space. A configuration is said to be \emph{generic} if $%
\Omega $ has maximum rank, $4\Delta -4$. Configurations of lower rank are
\emph{degenerate}; they have additional gauge symmetries and, consequently,
fewer degrees of freedom. For example, any ``vacuum'' solution $F^{K}=0$ is
maximally degenerate since the symplectic form vanishes on it and hence no
local excitations can propagate around these configurations. Since the rank
cannot change under small deformations, generic sectors form open sets in
field space, whereas degenerate configurations define sets of measure zero.

\emph{Regular} configurations are those on which the constraints $C_{K}=0$
are functionally independent, that is, they define a smooth surface with a
unique tangent space on an open set around the configuration.\footnote{%
Regular configurations also form open sets in field space, while irregular
ones form sets of measure zero.} This means that the variations
\begin{equation}
\delta C_{K}=\frac{1}{2}\,\Omega _{KL}^{ij}\,\delta F_{ij}^{L}\,,  \label{Ck}
\end{equation}%
at a regular configuration, are $\Delta $ linearly independent vectors in
phase space. Consequently, regularity is satisfied if the Jacobian of Eq. (%
\ref{Ck}) has maximal rank, $\Delta $, where now $\Omega _{KL}^{ij}$ has to
be regarded as a $\Delta \times 6\Delta $ matrix with indices $(_{K})$ and $%
\left( _{L}^{ij}\right) $.

Note that genericity does not imply regularity, and vice-versa. This is
because in one case the components $\Omega$ are regarded as the entries of a
square matrix and, in the other, as the entries of a rectangular one.

Dirac's Hamiltonian formalism requires that, in an open set, the symplectic
matrix be of constant rank and the constraints be functionally independent.
Hence, we call \emph{canonical sectors} of phase space those that are
simultaneously generic and regular, because that is where the canonical
formalism applies without modifications. Around degenerate or irregular
configurations, the dynamical content of the theory requires extending
Dirac's formalism as in \cite{Saavedra-Troncoso-Zanelli,Miskovic-Zanelli}.

\section{Canonical sectors}

The identification of the canonical sectors for a Chern-Simons theory in
general is a nontrivial task. However, a little bit of information about the
group and the invariant tensor allows, in some cases, to identify these
sectors. Let us split the generators as $\mathbf{G}_{K}=\{\mathbf{G}_{\bar{K}%
},\mathbf{Z}\}$. If the group admits a third rank invariant tensor $g_{KLM}$
such that $g_{\bar{K}\bar{L}z}:=g_{\bar{K}\bar{L}} $ is invertible, and $g_{%
\bar{K}zz}=0$, then the search for canonical sectors if much simpler. These
conditions are trivially fulfilled by a non-Abelian group of the form $G=%
\tilde{G}\otimes U(1)$, and also apply to a larger class of theories
including supergravity, for which the relevant group is super AdS$_{5}$, $%
SU(2,2|N)$.

Thus, the symplectic matrix reads
\begin{equation}
\Omega _{KL}^{ij}=\left(
\begin{array}{cc}
\Omega _{\bar{K}\bar{L}}^{ij} & \Omega _{\bar{K}z}^{ij} \\
\Omega _{\bar{L}z}^{ij} & \Omega _{zz}^{ij}%
\end{array}
\right) =-3k\,\varepsilon ^{ijkl}\left(
\begin{array}{cc}
g_{\bar{K}\bar{L}\bar{M}}\,F_{kl}^{\bar{M}}+g_{\bar{K}\bar{L}}\,F_{kl}^{z} &
g_{\bar{K}\bar{M}}\,F_{kl}^{\bar{M}} \\
g_{\bar{L}\bar{M}}\,F_{kl}^{\bar{M}} & \alpha F_{kl}^{z}%
\end{array}
\right) \,,  \label{Omega}
\end{equation}
where $\alpha :=g_{zzz}$.

Consider the following class of configurations,
\begin{equation}
\Omega _{\bar{K}\bar{L}}^{ij}=\mbox{non-degenerate,}\qquad \det
(F_{ij}^{z})\neq 0\,.  \label{class}
\end{equation}
For this kind of configurations, the rank $\Re (\Omega _{KL})=\Re (\Omega _{%
\bar{K}\bar{L}})=4\Delta -4$, and therefore, they provide generic
backgrounds.\footnote{%
This can be explicitly seen from $\Re \left( \Omega _{KL}^{ij}\right) =\Re
\left( \Omega _{\bar{K}\bar{L}}^{ij}\right) +\Re \left( \Sigma ^{ij}\right) $
, where the matrix $\Sigma ^{ij}$ vanishes for $C_{M}\approx 0,$ as it is
given by $\Sigma ^{ij}=\left( \delta _{k}^{i}C_{z}-\Omega _{z\bar{K}%
}^{il}\left( \Omega ^{-1}\right) _{lk}^{\bar{K}\bar{L}}C_{\bar{L}}\right)
\varepsilon ^{jknm}F_{nm}^{z}\,$, and $\Omega ^{-1}$ is the inverse of the
invertible block $\Omega _{\bar{K}\bar{L}}^{ij}$ only.}

Among the configurations (\ref{class}), the regular ones are those for which
the variations of the constraints (\ref{Ck}) are linearly independent, and
this depends on the value of $\alpha $,
\begin{equation}
\delta C_{\bar{K}}=\frac{1}{2}\,\Omega _{\bar{K}\bar{L}}^{ij}\,\delta
F_{ij}^{\bar{L}}+\frac{1}{2}\,\Omega _{\bar{K}z}^{ij}\,\delta
F_{ij}^{z}\,,\qquad \delta C_{z}=\frac{1}{2}\,\Omega _{\bar{K}%
z}^{ij}\,\delta F_{ij}^{\bar{K}}-\frac{3}{2}\,k\alpha \,\varepsilon
^{ijkl}F_{ij}^{z}\,\delta F_{kl}^{z}\,.  \label{delta C}
\end{equation}

If the $(\Delta -1)\times 6(\Delta -1)$ block $\Omega _{\bar{K}\bar{L}}^{ij}$
is non-degenerate, $\delta C_{\bar{K}}$ represent $\Delta -1$ linearly
independent vectors expressed as a linear combination of $\delta F^{\bar{K}}$%
. The problem of regularity then, reduces to the question of linear
independence of the vector $\delta C_{z}$ relative to the $\delta C_{K}$'s.

If $\alpha\neq 0$, the block $\Omega _{zz}$ is non-vanishing, so that $%
\delta C_{z}$ in (\ref{delta C}) always contains the term proportional to $%
\delta F^{z}$, giving a vector linearly independent from $\delta C_{\bar{K}%
}. $ Therefore, one concludes that \medskip

\emph{For }$\alpha\neq 0$\emph{, the dynamics of the sectors defined by open
sets around configurations of the form }(\ref{class})\emph{\ is always
canonical.} \medskip

However, for $\alpha =0$, the variations of $C_{z}$ do not depend on $F^{z}$
but on the remaining components $F^{\bar{K}}$. In particular, note that for
a configuration with $F^{\bar{K}}=0$, the block $\Omega_{\bar{K}\bar{L}%
}^{ij} =-3k\,\varepsilon ^{ijkl}g_{\bar{K}\bar{L} }\,F_{kl}^{z}$ is
invertible and hence, this configuration is generic. However, this
configuration is irregular because $\delta C_{z}=0$. One therefore concludes
that,\medskip

\emph{For} $\alpha =0$, \emph{the theory contains additional irregular
sectors which do not exist if} $\alpha \neq 0$. \emph{Thus, a necessary
condition to ensure regularity for configurations of the form} (\ref{class})%
\emph{,} \emph{for} $\alpha =0$\emph{, is that at least one component of} $%
F^{\bar{K}}$ \emph{does not vanish.}\medskip

In a canonical sector the counting of degrees of freedom can be safely done
following the standard procedure \cite{Dirac}, and in this case the number
is $N=\Delta-2$ \cite{Banados-Garay-Henneaux}.

A simple example of a solution of the constraints in the canonical sector is
one whose only nonvanishing components of $F^{\bar{K}}$ is
\begin{equation}
F_{12}^{\bar{K}}\,dx^{1}\wedge dx^{2}\neq 0\,,
\end{equation}
for at least one $\bar{K}$ and
\begin{equation}
F_{34}^{z}=0\,,\qquad \mbox{with  }\qquad \det \left( F_{ij}^{z}\right) \neq
0\,.  \label{explicit solution}
\end{equation}

\section{Constraints and charge algebra}

The canonical sectors satisfy all conditions necessary for the Dirac
formalism to apply. That is, the first and second class constraints can be
clearly defined and the counting and identifying of degrees of freedom can
be explicitly done \cite{Henneaux:1990au}. The explicit separation between
first and second class constraints might be extremely difficult or
impossible to carry out. This obstacle prevents, among other things, the
canonical construction of the conserved charges.

The advantage of the class of canonical sectors described above, is that
this splitting can be performed explicitly and, as a consequence, the
conserved charges and their algebra are obtained following the
Regge-Teitelboim approach \cite{Regge-Teitelboim}.

\subsection{Hamiltonian structure}

The Hamiltonian formalism applied to CS systems was performed in \cite%
{Banados-Garay-Henneaux} and can be easily extended to the supersymmetric
case \cite{Chandia-Troncoso-Zanelli}. The action in Hamiltonian form reads

\begin{equation}
I\left[ A\right] =\int d^{5}x\,\left( \mathcal{L}_{K}^{i}\,\dot{A}%
_{i}^{K}-A_{0}^{K}C_{K}\right) \,,
\end{equation}
where the constraints $C_{K}$ are defined in (\ref{Ck}),

\begin{equation}
\mathcal{L}_{K}^{i}\equiv k\,\varepsilon ^{ijkl}\,g_{KLM}\left(
F_{jk}^{L}A_{l}^{M}-\frac{1}{4}\,f_{NS}^{\quad
\;M}A_{j}^{L}A_{k}^{N}A_{l}^{S}\right) \,,  \label{linear in V}
\end{equation}
and $f_{NS}^{\quad \;M}$ are the structure constants of the Lie group.
Additional constraints arise from the definition of the momenta,
\begin{equation}
\phi _{K}^{i}=\pi _{K}^{i}-\mathcal{L}_{K}^{i}\approx 0\,,
\end{equation}
and they satisfy
\begin{equation}
\left\{ \phi _{K}^{i},\phi _{L}^{j}\right\} =\Omega _{KL}^{ij}\,.
\label{algebra3}
\end{equation}
Since the symplectic matrix $\Omega _{KL}^{ij}$ has at least four null
eigenvectors, some $\phi $'s are first class and the explicit separation
cannot be performed in general. However, for the class of canonical
configurations described in the previous chapter, it is possible to separate
them as
\begin{equation}
\begin{array}{ll}
\mbox{First class}:\qquad & G_{K}=-C_{K}+\nabla _{i}\phi _{K}^{i}\approx
0\,,\medskip \\
& \mathcal{H}_{i}=F_{ij}^{z}\left( \phi _{z}^{j}-\phi _{\bar{K}}^{k}\left(
\Omega ^{-1}\right) _{kl}^{\bar{K}\bar{L}}\Omega _{\bar{L}z}^{lj}\right)
=F_{ij}^{K}\phi _{K}^{j}\approx 0\,,\medskip \\
\mbox{Second class}:\qquad & \phi _{\bar{K}}^{i}\approx 0\,,%
\end{array}
\label{1st+2nd}
\end{equation}
where $\nabla _{i}\phi _{K}^{i}=\partial _{i}\phi _{K}^{i}+f_{KL}^{\quad
\;M}A_{i}^{L}\phi _{M}^{i}\,,$ is the covariant derivative, and the
constraints $G_{K}$ satisfy the algebra
\begin{equation}
\left\{ G_{K},G_{L}\right\} =f_{KL}^{\quad \;M}\,G_{M}\,,\qquad \qquad
\left\{ G_{K},\phi _{L}^{i}\right\} =f_{KL}^{\quad \;M}\,\phi _{M}^{i}\,.
\end{equation}

The constraints $G_{K}$ and $\mathcal{H}_{i}$ are generators of gauge
transformations and improved spatial diffeomorphisms, respectively.\footnote{%
An improved diffeomorphism, $\delta _{\xi }A_{\mu }^{K}=F_{\mu \nu
}^{K}\,\xi ^{\nu }$, differs from the Lie derivative by the gauge
transformation with $\lambda ^{K}=-\xi ^{\mu }A_{\mu }^{K}$\thinspace .} The
second class constraints can be eliminated by introducing Dirac brackets,
\begin{equation}
\left\{ X,Y\right\} ^{*}\equiv \left\{ X,Y\right\} -\int\limits_{\Sigma
}d^{4}x\,\left\{ X,\phi _{\bar{K}}^{i}\left( x\right) \right\} \left( \Omega
^{-1}\right) _{ij}^{\bar{K}\bar{L}}\left( x\right) \left\{ \phi _{\bar{L}%
}^{j}\left( x\right) ,Y\right\} \,,
\end{equation}
and the reduced phase space is parametrized by $\left\{ A_{i}^{\bar{K}%
},A_{i}^{z},\pi _{z}^{i}\right\} $.

\subsection{Conserved charges}

The separation (\ref{1st+2nd}) allows the construction of the conserved
charges and the algebra following the Regge-Teitelboim approach \cite%
{Regge-Teitelboim}. The symmetry generators are
\begin{equation}
G_{Q}\left[ \lambda \right] =G\left[ \lambda \right] +Q\left[ \lambda \right]
\,,
\end{equation}
where
\begin{equation}
G\left[ \lambda \right] =\int\limits_{\Sigma }d^{4}x\,\lambda ^{K}G_{K}\,,
\end{equation}
and $Q\left[ \lambda \right] $ is a boundary term such that $G_{Q}\left[
\lambda \right] $ has well-defined functional derivatives. According to the
Brown-Henneaux theorem, in general the charge algebra is a central extension
of the gauge algebra \cite{Brown-HenneauxQ},
\begin{equation}
\left\{ Q\left[ \lambda \right] ,Q\left[ \eta \right] \right\}^* =Q\left[ %
\left[ \lambda ,\eta \right] \right] +C\left[ \lambda ,\eta \right] \,,
\label{charge algebra}
\end{equation}
where $\left[ \lambda ,\eta \right] ^{K}=f_{LM}^{\quad \;K}\lambda ^{L}\eta
^{M}$.

Thus, being in a canonical sector allows writing the charges as (see
Appendix),\footnote{%
Hereafter, the forms are defined at the spatial section $\Sigma $.}
\begin{equation}
Q\left[ \lambda \right] =-3k\int\limits_{\partial \Sigma }g_{KLM}\,\lambda
^{K}\bar{F}^{L}A^{M},  \label{charge}
\end{equation}%
where $\bar{F}$ is the background field strength and the parameters $\lambda
^{K}(x)$ approach covariantly constant fields at the boundary, and the
central charge is
\begin{equation}
C\left[ \lambda ,\eta \right] =3k\int\limits_{\partial \Sigma
}g_{KLM}\,\lambda ^{K}\bar{F}^{L}d\eta ^{M}.  \label{central charge}
\end{equation}

The charge algebra can be recognized as the WZW$_{4}$ extension of the full
gauge group \cite{Losev-Moore-Nekrasov-Shatashvili}. In an irregular sector
the charges are not well defined and the naive application of the Dirac
formalism would at best lead to a charge algebra associated to a subgroup of
$G$.

\section{Discussion}

Configurations in the canonical sectors satisfy all necessary conditions to
safely apply the Dirac formalism to five-dimensional CS\ theories. The
identification of these sectors in CS theories considered here, allows the
explicit separation of first and second class constraints. Consequently, the
conserved charges and their algebra are constructed following the
Regge-Teitelboim approach.

As a direct application of these results in the context of supergravity,
canonical BPS states saturating a Bogomol'nyi bound for the conserved
charges (\ref{charge}) can be constructed \cite{Ms-T-Z}. One should expect
that these results extend to higher odd dimensions. Indeed, conserved
charges have been found in the purely gravitational case following a
different approach \cite{MOTZ}.

Overlooking regularity obstructs the construction of well defined canonical
generators. Consider theories with $\alpha =0$ which accept simple generic
configurations of the form (\ref{class}), given by $F^{\bar{K}}=0$ with $%
\det \left( F^{z}\right) \neq 0$. These configurations are generic but
irregular (and therefore not canonical), since $\delta C_{z}$ in Eq. (\ref%
{delta C}) identically vanishes. If one naively chooses a configuration of
this type as a background in the expression for the charges (\ref{charge}),
one obtains that the $U(1)$ charge identically vanishes. This example simply
reflects the fact that, for irregular configurations, $U(1)$ generator
becomes functionally dependent, so that the naive application of Dirac's
formalism within irregular sectors would lead to ill-defined expression for
the charges.

Canonical sectors represent typical initial configurations around which one
can prepare the system to let it evolve. In its evolution, the system may
reach degenerate or irregular configurations. Although either degenerate or
irregular configurations are easily identified, it is not yet fully
understood how to deal with the dynamics around them in general, and it is
clear that one must proceed with caution. However, the analysis of
degenerate mechanical systems provide simple models that describe
irreversible processes in which a system may evolve into a configuration
with fewer degrees of freedom \cite{Saavedra-Troncoso-Zanelli}. From these
results one infers that, under certain conditions, a CS system may fall into
a degenerate state from which it can never escape, losing degrees of freedom
in an irreversible manner \cite{Saavedra-Troncoso-Zanelli}; or it can also
pass through irregular states unharmed \cite{Miskovic-Zanelli}.

The possibility that the fate of an initial configuration in a canonical
sector of a higher dimensional CS system could be to end in a degenerate
state, leads to an interesting effect: In Ref. \cite%
{Hassaine-Troncoso-Zanelli}, a new kind of eleven-dimensional supergravity
was constructed as a CS system for the M-algebra. It was observed that the
theory admits a class of vacuum solutions of the form $S^{10-d}\times
Y_{d+1} $, where $Y_{d+1}$ is a warped product of $R$ with a $d $%
-dimensional spacetime. Remarkably, it was found that a nontrivial
propagator for the graviton exists only for $d=4$ and positive cosmological
constant, and that perturbations of the metric around this solution
reproduce linearized General Relativity around four-dimensional de Sitter
spacetime.

Since this solution is a degenerate state one may regard it as the final
stage of a wide class of initial canonical configurations that underwent a
sort of dynamical dimensional reduction.

\section*{Acknowledgments}

We gratefully acknowledge insightful discussions with M. Henneaux. This work
was partially funded by FONDECYT grants 1020629, 1040921, 1051056, and
3040026. The generous support to CECS by Empresas CMPC is also acknowledged.
CECS is a Millennium Science Institute and is funded in part by grants from
Fundaci\'{o}n Andes and the Tinker Foundation.

\section*{Appendix}

On the reduced phase space $\phi_{\bar{K}}^{i}=0$, the generators are
\begin{equation}
G_{\bar{K}}=-C_{\bar{K}}\,,\qquad \qquad G_{z}=-C_{z}+\partial _{i}\phi
_{z}^{i}\,,
\end{equation}
and the smeared generators become
\begin{equation}
G\left[ \lambda \right] =3k\int\limits_{\Sigma }\left\langle \mathbf{\lambda
F}^{2}\right\rangle +\int\limits_{\Sigma }\left\langle \lambda ^{z}d\mathbf{%
\Phi }\right\rangle \,,  \label{generators}
\end{equation}
where the 3-form $\mathbf{\Phi }=\mathbf{\pi }-k\mathbf{\,}\left( \left\{
\mathbf{\mathbf{A,F}}\right\} -\frac{1}{2}\mathbf{\,\mathbf{A}}^{3}\right) $
is dual of the constraints, \emph{i. e.}, $\mathbf{\Phi }_{jkl}=\phi
_{z}^{i}\,\varepsilon _{ijkl}$.

The variation of the generators (\ref{generators}) yields
\begin{equation}
\delta G\left[ \lambda \right] =6k\int\limits_{\Sigma }\left\langle \mathbf{%
\lambda \mathbf{F}}\nabla \delta \mathbf{A}\right\rangle
+\int\limits_{\Sigma }\left\langle \lambda ^{z}d\delta \mathbf{\Phi }%
\right\rangle \,.  \label{varied G}
\end{equation}%
Integrating by parts this expression in order to obtain a well defined
functional derivative, the leftover boundary term is the variation of the
charge
\begin{equation}
\delta Q\left[ \lambda \right] =-6k\int\limits_{\partial \Sigma
}\left\langle \mathbf{\lambda F}\delta \mathbf{A}\right\rangle
-2k\int\limits_{\partial \Sigma }\left\langle d\lambda ^{z}\mathbf{A}\delta
\mathbf{A}\right\rangle -\int\limits_{\partial \Sigma }\left\langle \lambda
^{z}\delta \mathbf{\Phi }\right\rangle \,.  \label{delta Q}
\end{equation}%
This expression can be integrated out provided the connection is fixed at
the boundary, as
\begin{equation}
\mathbf{A\;}\longrightarrow \mathbf{\;\bar{A}}\qquad \mbox{at}\mathbf{\;}%
\partial \Sigma \,,  \label{ac}
\end{equation}%
where $\mathbf{\bar{A}}$ is a background configuration in the canonical
sector. Then, the charges are
\begin{equation}
Q\left[ \lambda \right] =-6k\int\limits_{\partial \Sigma }\left\langle
\mathbf{\lambda \bar{F}A}\right\rangle \,-2k\int\limits_{\partial \Sigma
}\left\langle d\lambda ^{z}\mathbf{\bar{A}A}\right\rangle \,,  \label{Q}
\end{equation}%
where the term proportional to $\mathbf{\Phi }$ vanishes on the constraint
surface.

Note that the second term $\left\langle d\lambda ^{z}\mathbf{\bar{A}A}%
\right\rangle$ identically vanishes when it is evaluated on the background $%
\mathbf{A}= \mathbf{\bar{A}}$, since $g_{zKL}\,\bar{A}^{K}\bar{A}^{L}\equiv
0 $.

Since the asymptotic behavior of the fields approaches a background
configuration as $\mathbf{A}\sim \mathbf{\bar{A}}+\Delta \mathbf{A}$, the
parameters of the asymptotic symmetries are of the form $\lambda \sim \bar{%
\lambda}+\Delta \lambda $. In particular, $\bar{\lambda}^{z}$ is a constant,
and then
\begin{equation}
\left\langle d\lambda ^{z}\mathbf{\bar{A}\mathbf{A}}\right\rangle \sim
\left\langle d\left( \Delta \lambda ^{z}\right) \mathbf{\bar{A}}\left(
\Delta \mathbf{\mathbf{A}}\right) \right\rangle \,,
\end{equation}
is a subleading contribution which vanishes as it approaches the boundary.
Therefore, the charges are given by (\ref{charge})
\begin{equation}
Q\left[ \lambda \right] =-6k\int\limits_{\partial \Sigma }\left\langle
\mathbf{\lambda \bar{F}A}\right\rangle \,.  \label{bar Q}
\end{equation}


\begin{thebibliography}{99}
\bibitem{Achucarro-Townsend} A. Ach\'{u}carro and P. K. Townsend, \emph{%
Phys. Lett.} \textbf{B180} (1986) 89.
%%CITATION = PHLTA,B180,89;%%

\bibitem{Witten'88} E. Witten, \emph{Nucl. Phys.} \textbf{B311} (1988) 46.
%%CITATION = NUPHA,B311,46;%%

\bibitem{Chamseddine1} A. Chamseddine, \emph{Phys. Lett.} \textbf{B233}
(1989) 291.
%%CITATION = PHLTA,B233,291;%%

\bibitem{Troncoso-Zanelli'00} R. Troncoso and J. Zanelli, \emph{Class.
Quant. Grav.} \textbf{17} (2000) 4451.
%%CITATION = HEP-TH 9907109;%%

\bibitem{Chamseddine2} A. Chamseddine, \emph{Nucl. Phys.} \textbf{B346}
(1990) 213.
%%CITATION = PHLTA,B228,75;%%

\bibitem{Troncoso-Zanelli'98} R. Troncoso and J. Zanelli, \emph{Phys. Rev.}
\textbf{D58}: R101703 (1998).
%%CITATION = HEP-TH 9710180;%%

\bibitem{Troncoso-Zanelli'99} R. Troncoso and J. Zanelli, \emph{Int. J.
Theor. Phys.} \textbf{38} (1999) 1181.
%%CITATION = HEP-TH 9807029;%%

\bibitem{Troncoso-Zanelli'99a} R. Troncoso and J. Zanelli, ``Chern-Simons
Supergravities with Off-Shell Local Superalgebras'', in \emph{Black Holes
and Structre of the Universe}, editors C. Teitelboim and J. Zanelli (World
Scientific, Singapore, 1999). Archive: \texttt{hep-th/9902003}.
%%CITATION = HEP-TH 9902003;%%

\bibitem{Banados-Troncoso-Zanelli} M. Ba\~{n}ados, R. Troncoso and J.
Zanelli, \emph{Phys. Rev.} \textbf{D54 }(1996) 2605.
%%CITATION = GR-QC 9601003;%%

\bibitem{HOT} M. Hassaine, R. Olea and R. Troncoso, \emph{Phys. Lett.}
\textbf{B599} (2004) 111.
%%CITATION = HEP-TH 0210116;%%

\bibitem{Hassaine-Troncoso-Zanelli} M. Hassaine, R. Troncoso and J. Zanelli,
\emph{Phys. Lett.} \textbf{B596 }(2004) 132; \emph{Proc. Sci.} \textbf{%
WC2004} (2005) 006.
%%CITATION = HEP-TH 0306258;%%
%%CITATION = HEP-TH 0503220;%%

\bibitem{Floreanini-Percacci} R. Floreanini and R. Percacci,\ \emph{Phys.
Lett.} \textbf{B224} (1989) 291.
%%CITATION = PHLTA,B224,291;%%

\bibitem{Floreanini-Percacci-Rajaraman} R. Floreanini, R. Percacci and R.
Rajaraman, \emph{Phys. Lett.} \textbf{B231} (1989) 119.
%%CITATION = PHLTA,B231,119;%%

\bibitem{Nair-Schiff} V. P. Nair and J. Schiff, \emph{Phys. Lett.} \textbf{%
B246} (1990) 423; \emph{Nucl. Phys.} \textbf{B371} (1992) 329.
%%CITATION = NUPHA,B371,329;%%
%%CITATION = NUPHA,B371,329;%%

\bibitem{Banados-Garay-Henneaux} M. Ba\~{n}ados, L. J. Garay and M.
Henneaux, \emph{Phys. Rev.} \textbf{D53} (1996) 593; \emph{Nucl. Phys.}
\textbf{B476} (1996) 61.
%%CITATION = HEP-TH 9506187;%%
%%CITATION = HEP-TH 9605159;%%

\bibitem{Saavedra-Troncoso-Zanelli} J. Saavedra , R. Troncoso and J.
Zanelli, \emph{J. Math. Phys.} \textbf{42} (2001) 4383; \emph{Rev. Mex. Fis.}
\textbf{48} (2002) SUPPL 387-89.
%%CITATION = HEP-TH 0011231;%%
%%CITATION = RMXFA,48,SUPPL387;%%

\bibitem{Henneaux-Teitelboim-book} M. Henneaux and C. Teitelboim,
``Quantization of Gauge Systems'', Univ. Pr. (1992) 520 p., Princeton, USA.

\bibitem{Miskovic-Zanelli} O. Mi\v{s}kovi\'{c} and J. Zanelli, \emph{J.
Math. Phys.} \textbf{44} (2003) 3876; ``Irregular Hamiltonian Systems'',\
Proceedings of the XIII Chilean Symposium of Physics (Concepci\'{o}n, Chile,
2002). Archive:\ \texttt{hep-th/0301256}.
%%CITATION = HEP-TH 0301256;%%
%%CITATION = HEP-TH 0302033;%%

\bibitem{Chandia-Troncoso-Zanelli} O. Chand\'{\i}a, R. Troncoso and J.
Zanelli, ``Dynamical Content of Chern-Simons Supergravity'', in \emph{Trends
in Theoretical Physics II}, H. Falomir, R .E. Gamboa Saravi and F.
Schapopsnik, Eds. AIP Conf. Proceedings 484, 231-237, Amer. Inst. Phys.,
Woodbury, NY, 1999. Archive: \texttt{hep-th/9903204}.
%%CITATION = HEP-TH 9903204;%%

\bibitem{Miskovic} O. Mi\v{s}kovi\'{c}, \textquotedblleft Dynamics of
Wess-Zumino-Witten and Chern-Simons theories\textquotedblright , Ph.D.
Thesis (Universidad de Santiago de Chile, Jan. 2004). Archive: \texttt{%
hep-th/0401185}.
%%CITATION = HEP-TH 0401185;%%

\bibitem{Buffenoir-Henneaux-Noui-Roche} E. Buffenoir, M. Henneaux, K. Noui
and Ph. Roche, \emph{Class. Quant. Grav.} \textbf{21} (2004) 5203.
%%CITATION = GR-QC 0404041;%%

\bibitem{Burgers} D. V. Choudnovsky and G. V. Choudnovsky, \emph{Nuovo Cim.}
\textbf{B40}, 339 (1977).

\bibitem{Landau-Lifschitz} Landau, L. D. and Lifschitz, \emph{E. M. Fluid
Mechanics}, 2nd ed. Oxford, England: Pergamon Press, 1982, Chapter XII.

\bibitem{Dirac} P. A. M. Dirac, Lectures on Quantum Mechanics, Yeshiva
University, New York, 1964.

\bibitem{Henneaux:1990au} M.~Henneaux, C.~Teitelboim and J.~Zanelli, Nucl.\
Phys.\ B \textbf{332}, 169 (1990).
%%CITATION = NUPHA,B332,169;%%

\bibitem{Regge-Teitelboim} T. Regge and C. Teitelboim, \emph{Ann. Phys.
(N.Y.)} \textbf{88} (1974) 286.
%%CITATION = APNYA,88,286;%%

\bibitem{Brown-HenneauxQ} J. D. Brown and M. Henneaux, \emph{J. Math. Phys.}
\textbf{27} (1986) 489; \emph{Commun. Math. Phys.} \textbf{104} (1986) 207.
%%CITATION = CMPHA,104,207;%%
%%CITATION = JMAPA,27,489;%%

\bibitem{Losev-Moore-Nekrasov-Shatashvili} A. Losev, G. Moore, N. Nekrasov
and S. Shatashvili, \emph{Nucl. Phys. Proc. Suppl.} \textbf{46} (1996) 130.
%%CITATION = HEP-TH 9509151;%%

\bibitem{Ms-T-Z} O. Mi\v{s}kovi\'{c}, R. Troncoso and J. Zanelli,
``Dynamics and BPS states of AdS$_{5}$ Supergravity with Gauss-Bonnet term".
Preprint CECS-PHY-05/07.

\bibitem{MOTZ} P. Mora, R. Olea, R. Troncoso and J. Zanelli, JHEP \textbf{%
0406} (2004) 036.
%%CITATION = HEP-TH 0405267;%%

\end{thebibliography}
\end{document}